\documentclass[%
 reprint,
superscriptaddress,
 amsmath,amssymb,
 aps,
prd,
]{revtex4-2}

\usepackage{dcolumn}
\usepackage[utf8]{inputenc}
\usepackage{graphicx,fancybox,float,comment,bm}
\usepackage[colorlinks=true, linkcolor=blue, citecolor=blue, urlcolor=blue]{hyperref}
\makeatother
\usepackage[dvipsnames]{xcolor}
\usepackage{amsmath,amssymb,amsfonts}

\begin{document}
\title{Anomalous resonance in Weyl semimetals: A holographic study of non-linear effects}
\author{Maximilian Gaschler}
\affiliation{Institute for Theoretical Physics and Astrophysics,
 Julius-Maximilians-Universität Würzburg, Am Hubland, 97074 Würzburg, Germany}
\email{maximilian.gaschler@uni-wuerzburg.de}

\author{Andreas Schäfer}
\affiliation{
Institute for Theoretical Physics, University of Regensburg, D-93040 Regensburg, Germany
}
\email{andreas.schaefer@physik.uni-regensburg.de}
\affiliation{Department of Physics, National Taiwan University, Taipei, Taiwan 106}

\author{Sebastian Waeber}
\affiliation{Department of Physics, Ben-Gurion University of the Negev,
David Ben Gurion Boulevard 1, Beer Sheva 84105, Israel}
\email{waeber@post.bgu.ac.il}

\begin{abstract}
We consider hairy, magnetic black brane solutions to Einstein-Maxwell-Chern-Simons-dilaton theory with full back-reaction of the scalar and gauge fields and compute the time evolution with time dependent sources. The dual field theory's 't Hooft anomaly leads to long-lived current oscillations in the low temperature limit long after the electric pulse which created these modes has been switched off. We find that, generally, non-linear effects increase the decay rates of long-lived modes by a temperature dependent amount. Nonetheless, our  results suggest that the life-time of time-translation symmetry breaking states can be made arbitrarily large if the temperature after the quench is sufficiently small. Experimentally this might be realizable in Weyl semimetals. 
\end{abstract}
\maketitle

\newcommand{\MG}[1]{{\textcolor{Blue}{ [MG:#1] }}}
\newcommand{\AS}[1]{{\textcolor{Green}{ [AS:#1] }}}
\newcommand{\SW}[1]{{\textcolor{Red}{ [SW:#1] }}}
\newcommand{\HL}[1]{{#1}}

\section{Introduction}

Over the last thirty years the AdS/CFT correspondence, in its weak formulation the duality of a conformal quantum field theory in $d$ dimensions and gravity in a $d+1$ dimensional anti-de Sitter spacetime  
\cite{Maldacena:1997re}, has proven to be a very powerful tool for many different research fields, ranging from condensed matter physics in low-dimensional systems to quantum gravity in extended space-times.  Within this vast range of physics applications, the fluid gravity duality provides an exceptionally useful perspective on problems in fluid dynamics by mapping them to gravity \cite{Bhattacharyya:2007vjd}.  In this context one of the most noteworthy advances  was to understand the important role  anomalies may play in hydrodynamics \cite{Erdmenger:2008rm,Banerjee:2008th,Landsteiner:2011cp,Son:2009tf,Jensen:2012kj,Golkar:2012kb,Jensen:2013kka,Jensen:2013rga}, which relied on, or were guided by computations in holography.
 They yield anomalous transport phenomena, such as the chiral vortical effect \cite{Kharzeev:2015znc} or the chiral magnetic effect \cite{Fukushima:2008xe}, and holography provides a powerful framework to study the effects of anomalies on dynamics of strongly coupled field theories. Through the holographic duality,  solutions to the Einstein-Maxwell-Chern-Simons theory are mapped to states  in a CFT coupled to a $U(1)$ gauge theory with a 't Hooft anomaly. \\\indent 
Related to the discussion of anomalous resonance \cite{Ammon:2017ded,Haack:2018ztx,Waeber:2024ilt,Meiring:2023wwi,Ammon:2020rvg} are  quantum time crystals \cite{Wilczek:2012jt}, i.e. systems which violate time translation symmetry and whose ground state shows a periodic time dependence. Soon after they were proposed, no-go theorems were discovered \cite{Watanabe:2014hea}, nonetheless, the ability to create driven time crystals, or construct approximate time crystals with long-lived excitations, or circumvent the no-go theorems in holographic theories, due to the semi-classical nature of the large $N$ limit, fostered a vast field of research that we can only partially summarize here.  It can be subdivided by distinguishing discrete and continuous time crystals (CTCs) where the latter became a lively field of research only very recently. Both lead to anomalous transport phenomena and can be studied using holography. While discrete time crystals (DTCs) require periodic driving, CTCs do not. They are autonomous systems that show auto-oscillations. DTCs were discovered first \cite{Choi:2017jba,Zhang:2017bwh,Yao:2017dwx,Else_2020} and CTC shortly afterwards  \cite{PhysRevA.99.053605,PhysRevX.15.011055,Kongkhambut:2022ngl,Liu:2024yze}.
In fact, CTCs and DTCs are closely related as can be seen, e.g., in \cite{Kongkhambut:2024zbs}. Weyl semimetals open many possibilities to find DTCs, which is especially interesting because of potential applications in information science, see e.g. \cite{Frey:2021iix}, and metrology. Experimentally  the Adler-Bell-Jackiw anomaly has been studied in Weyl materials, e.g. in \cite{Bevan_1997}. In superfluids, long-lived metastable states with periodic motion have also been already observed experimentally, see e.g. \cite{Autti:2021uen, Autti_2020}, the no-go theorem does not apply as the system is kept in an excited metastable state. 

The connection between time crystals and AdS/CFT and string theory is quite straight forward. Much work was invested in holographic calculations of the real and imaginary part of quasi-normal modes (QNMs), also by our group, see e.g. \cite{Waeber:2015oka,Waeber:2024ilt,Waeber:2018bea,Waeber:2018dae}, and time crystals, on a linearized level, can be regarded as QNMs with vanishing imaginary part of its frequency. Thus, many properties of time crystals can be obtained from the holographic calculation of QNMs. This motivates not only the present project, but also \cite{Wang:2025aoe}. 
A crucial point in all theoretical work on time crystals is that to be observable the time crystal must be robust and long lived, see e.g. \cite{Greilich:2023dcj}. In applications to condensed matter phenomena holographic calculations are typically based on models rather than fundamental theories and thus could be misleading for some details. Luckily there are examples of this type, see e.g. \cite{Greilich:2023dcj}.

\indent The authors of \cite{Wang:2025aoe} give a very nice example. They argue that the collective quantum dynamics of the superfluid \cite{Zhai_2007,PhysRevLett.118.080403}, Mott insulator \cite{Greiner_2002}, and CTC phase of a Bose-Einstein condensate embedded in a three-dimensional optical lattice can be modeled by the QNMs of an AdS-Reissner-Nordstr\"om black hole, and solve the corresponding Einstein equations. The result is a quantitative phase diagram containing these three phases. However, this calculation does not take back-reactions into account, which is our goal in the calculation presented here. The scope of this work is to explore subtle but relevant details of the anomalous resonance effect \cite{Ammon:2020rvg,Haack:2018ztx,Meiring:2023wwi,Waeber:2024ilt}, which might be observable in Weyl semimetals \cite{huang2015observation,zhang2016signatures,arnold2016negative,gooth2017experimental} using holographic techniques \cite{Landsteiner:2015lsa}. While resonances with back-reaction in similar set ups have been studied, e.g. in \cite{Ghosh:2021naw}, we also include back-reacting, dynamical scalar deformations in this work, while focusing on Weyl semimetal phenomenology, carefully examining the effect non-linear corrections have on long lived resonances in the small temperature limit. 

 The fact that our results appear to not depend on the specific matter content of the theory, hints at a broader validity, i.e. remarkable robustness. For the anomalous resonance, we consider a state in a CFT with anomalously broken $U(1)$ symmetry in an external magnetic field.  This state is driven out of equilibrium by, e.g., an external electric field quench. After all sources are turned off, the state relaxes freely. With  a sufficiently large Chern-Simons coupling, the decay rate of the induced current is exponentially small and approaches zero in the  low temperature limit in a linearized approximation \cite{Ammon:2017ded,Meiring:2023wwi,Waeber:2024ilt}.  Intriguingly, the long lifetime of these modes means that they might serve as experimentally measurable signatures of anomalous transport.  \\ \indent In this work, we argue that, for sufficiently small temperatures, it is necessary to solve the problem fully back-reacted. Linearized solutions suggest the existence of modes with infinite life-time in the zero-temperature limit. It is straightforward to see, even without any computation, that non-driven, permanently oscillating, exact solutions to Einstein equations might not exist in this set-up. Once higher-order effects are taken into account, the current oscillations will excite the dual metric. Those metric fluctuations should dissipate like regular quasi-normal modes, as the gravitational perturbations are not affected by the Chern-Simons term and without the anomaly even the previously non-dissipative modes start decaying once the Chern-Simons coupling is switched off. Thus, the current fluctuations could dissipate through higher order effects. Moreover, once the equations of motions are solved with full back-reaction, the initial electric field quench will increase the temperature, such that the final state oscillates around a $T\neq0$ state even when the initial state was at zero temperature. As the infinite life-time of the anomalous resonance modes in the linearized approximation crucially depends on the zero-temperature limit, their decay rate should become non-zero once we solve the bulk Einstein-Maxwell-Chern-Simons equations exactly.  We aim at quantifying this non-zero decay rate in the low temperature limit through a computation with full backreaction. 
\section{Holographic setup}
\label{S:MBB}
In holography, quantum field theories with a global anomaly are described by  Einstein-Maxwell-Chern-Simons theory in the bulk. In the following we work with a theory with three $U(1)$ currents with a 't Hooft anomaly dual to a gravity theory in five bulk dimensions. The additional symmetry allows us to find analytical equilibrium solutions (at least in the case with zero scalar field). The Einstein-Maxwell-Chern-Simons action deformed by an uncharged, real scalar field $\phi$, whose action is  denoted by $S_{matter}$, is given by
\begin{multline}
\label{E:ansatz}
	S = \frac{1}{16 \pi G_N}\int \sqrt{-g} \bigg( R + {12}  - \frac{1}{4} \delta_{ij} F^i{}_{\mu\nu}F^{j\,\mu\nu} + \\
    \lambda \epsilon^{\mu\nu\rho\sigma\tau} A^1{}_{\mu}F^2_{\nu\rho}F^3_{\sigma\tau} \bigg) d^5x\,+ S_{matter}\,.
\end{multline}
The first term, for the specific choice of $\lambda=1/4$ is the STU ansatz for the $D=5$, $N=2$ gauged supergravity action.  The matter action is of the form
\begin{equation}
\label{E:Smatter}
   S_{matter} = \frac{1}{16\pi G_N} \int \sqrt{-g} \Big(- (\nabla \phi)^2 - V(\phi)\Big).
\end{equation}
In the following we set the AdS radius $L$ to 1. The equations of motion resulting from \eqref{E:ansatz} are given by
\begin{align}
\begin{split}
\label{E:EOM}
	&R_{\mu\nu}-\frac{1}{2} Rg_{\mu\nu}-\partial_\mu\phi \partial_\nu \phi - (6  -\frac{V(\phi)+ \partial_\rho \phi\partial^\rho\phi}{2}) g_{\mu\nu}\\ & = \sum_i \left(\frac{1}{2} F^{i}{}_{\mu\alpha}F^{i}{}_{\,\nu}{}^{\alpha} - \frac{1}{8} g_{\mu\nu} F^{i}{}_{\alpha\beta} F^{i\,\alpha\beta}\right)  \\
	&\nabla_{\mu}F_{1}{}^{\mu\nu} = -\lambda \epsilon^{\nu\alpha\beta\gamma\delta}F_{2\,\alpha\beta}F_{3\,\gamma\delta}\\&
      \nabla^2 \phi =\frac{1}{2} V'(\phi)
\end{split}
\end{align}
and even permutations of the second equation for $F_{2\,\mu\nu}$ and $F_{3\,\mu\nu}$. The action \eqref{E:ansatz} possesses an $SO(3)_f$ flavor symmetry rotating the three $U(1)$ gauge fields which is broken by the Chern-Simons term. The ansatz
\begin{align}
\begin{split}
\label{E:ansatzsolution}
	ds^2 = -  A(r)dt^2 + \Sigma(r)^2 \left(dx^2+dy^2+dz^2 \right) + 2 dt dr\,, \\ \phi=\phi(r)\,,
	\;
	A^1 =  {B\, y} dz\,,
	\;
	A^2 = {B\, z} dx\,,
	\;
	A^3 = {B\, x} dy\,.
\end{split}
\end{align}
for a static solution preserves a diagonal $SO(3)_B$ subgroup of the $SO(3)_f$ symmetry and spatial rotations.  The ansatz \eqref{E:ansatzsolution}  is invariant under a simultaneous spatial rotation in the $x$, $y$, $z$ directions, and a rotation of the three $U(1)$ fields. It is also invariant under parity transformations $x^i\rightarrow-x^i$ of the spatial coordinates.\\ \indent \HL{As we are interested in a backreacted treatment, we consider in (\ref{E:ansatz}) a simplified  holographic model of a strongly coupled Weyl semimetal, which still captures the relevant physics of the anomalous resonance effect.   The main differences to standard holographic models for Weyl semimetals \cite{Landsteiner:2015lsa} are the presence of three $U(1)$ gauge fields as opposed to only one in \cite{Landsteiner:2015lsa} and a real, uncharged scalar field, compared to a charged scalar in \cite{Landsteiner:2015lsa}, the bulk dual of the chiral symmetry breaking operator. As discussed in \cite{Waeber:2024ilt}, the anomalous resonance effect is present for both theories with three $U(1)$ gauge fields and one  $U(1)$ gauge field, with identical linearized equations of motion for the respective gauge field perturbations. Thus, this detail is not expected to affect the results. The additional symmetry of theories with three $U(1)$ currents simplifies the computation in the absence of scalar fields. Including a complex scalar field coupled to the axial gauge fields is left for future work. Having also charged scalar fields goes beyond a small technicality, as  the constant magnetic field would introduce terms in the equations of motions for the scalar field that are incompatible with symmetric (spatial) boundary conditions, which are used in our numerical treatment later. Performing our computations and analysis both with and without a scalar deformation (\ref{E:Smatter}) of the Einstein-Maxwell-Chern-Simons theory serves as a test of robustness of our results.   }
\\ \indent
Our goal is to find fully back-reacted, dynamical solutions to this theory, corresponding to a thermal state that was driven out of equilibrium by an electric quench and track the response of the dual current. At low temperatures and sufficiently large Chern-Simons coupling a linearized analysis yielded decay rates of current oscillations that approached zero in the zero temperature limit, as shown in \cite{Meiring:2023wwi,Haack:2018ztx,Waeber:2024ilt,Ammon:2020rvg}. Naively this suggests the existence of a stable time-periodic state in the dual CFT. This time-translation invariance breaking state would not require external driving to be maintained. While the large $N$ limit for holographic theories implies that "holographic time crystals" (i.e. non-dissipative, stable, time-periodic states in the dual CFT, breaking the continuous time-translation invariance to a discrete subgroup) can exist, exact solutions with full back-reaction of the bulk gauge fields  are necessary to determine the exact decay rates of the current oscillations. We find that after taking into account non-linear effects, the state whose linearized treatment suggested infinitely long-lived current oscillations at zero temperature, is only an "approximate" time crystal: the decay rate of the time-translation invariance breaking oscillations can be made arbitrarily small by tuning the parameters such that the temperature after the quench $T_{final}$ is sufficiently small. However, in order to create this state, energy, in this case in form of a short-lived electric pulse, is injected, increasing the temperature, such that $T_{final}$ is never exactly zero. \\ \indent 
We choose the following ansatz for the metric and gauge field, protected by a residual discrete symmetry, composed of a rotation by $\pi/2$ in $SO(3)_B$ and a parity transformation: 
\begin{align}
\label{ansatz_1}
    ds^2 &= - A(t,u)\,dt^2-\frac{2 du dt}{u^2}+ \frac{s(t,u)^2}{u^2}\nonumber\\&\times\Big(q(t,u)\,dx^2+q(t,u)\,dy^2 +1/q(t,u)^2 dz^2 \Big)\\
    A^1 &= a^1_y(t,u) dy + B y dz\,,\,\,
	A^2 = a^2_x(t,u) dx + B z dx\,,\nonumber \\
	\,\,
	A^3 &= a^3_t(t,u) dt + B x dy\,,\,\,
   \phi=\phi(t,u),
   \label{ansatz_3}
\end{align}
with $a^1_y = a^2_x$. The coordinate $u$ is the inverted radial coordinate $u=1/r$. The Maxwell equations' constraint equation, corresponding to the charge conservation equation on the boundary, relates $a^3_t$ to $a^2_x$ (see the discussion in the Appendix \ref{numerics}).
We choose the potential of the scalar to be  
\begin{equation}
    V(\Phi)= -3 \phi^2 + \sigma \phi^4,
\end{equation}
with $\sigma>0$ corresponding to a bounded potential of a relevant scalar deformation of the boundary CFT.  The mass squared of the bulk scalar field being $-3$ yields integer powers for the leading near boundary behavior of the scalar field, introducing additional log-terms (in addition to the log-terms introduced through the gauge fields).
\\ \indent The near boundary expansion of the Einstein-Maxwell-Chern-Simons-dilaton equations yields the following near boundary behavior of the bulk fields in (\ref{ansatz_1})-(\ref{ansatz_3})
\begin{align}
\label{exp_start}
\phi(t,u)&=p_1u+\frac{1}{3} u^3 \left(3 p_1^3 \sigma +p_1^3\right) \log (u)\nonumber\\ &+u^3 p_2(t) + \mathcal{O}(u^4 \log(u))\\
a^2_x(t,u)&= a_0(t)+ u a_0'(t)  + u^2 a_2(t) \nonumber\\&+
 \frac{u^2}{2} \log(u) a_0''(t)  + \mathcal{O}(u^3 \log(u)),\\
 s(t,u) &=1-\frac{p_1^2 u^2}{6}+\frac{1}{18} u^4 \left(-3 p_1^4 \sigma -p_1^4\right) \log
   (u) \nonumber\\&+\mathcal{O}(u^4)
\end{align}
\begin{align}
   q(t,u) &= 1 + u^4 q_4(t) + \frac{u^4}{12} \log(u) a_0'^2 +\mathcal{O}(u^5 \log(u)),\\
   A(t,u) &= \frac{1}{u^2}-\frac{p_1^2}{3} + u^2 f_2(t) + 
 \frac{u^2}{18} \log(u) (9 B^2 - 2 p_1^4 \nonumber \\&- 6 p_1^4 \sigma + 6 a_0'(t)^2) +\mathcal{O}(u^3 \log(u)),
 \label{exp_end}
\end{align}

where, up to constants, $p_1$ is the source $\Lambda$ of the scalar operator $\mathcal{O}$, $p_2$ is its expectation value $\langle \mathcal{O}\rangle$, $a_0$ is the external boundary gauge field, sourcing the boundary current encoded in $a_2$,  and $f_2$ and $q_4$ determine the energy density and the pressure anisotropy respectively. For the expressions (\ref{exp_start}-\ref{exp_end}) we set the shift parameter $l$ of the radial shift freedom $u\rightarrow u/(1+l u)$ to zero. We choose a renormalization scheme such that the exact relations between the near boundary expansion coefficients $a_0$, $a_2$ and the boundary current $(J^2)^\mu$ is given by 
\begin{equation}
\label{cov_current}
    8 \pi G_5 \langle (J^2)^\mu\rangle = \delta^\mu_x r_h^3 (a_2(t)-\frac{3}{8} \partial_t^2 a^{(0)}),
\end{equation}
where $r_h$ is the radial position of the horizon. Note that we defined the covariant current in (\ref{cov_current}), as opposed to the consistent current (as done, e.g., in \cite{Landsteiner:2015lsa}). However, at late times, once sources corresponding to the electric pulse have been switched off, both consistent and covariant current are identical. All of our phenomenological analysis focuses on these late time modes.  In order to be able to make contact with experimental results for Weyl semimetals in future, we will express our final results in section \ref{S:numerics}  in more accessible units (i.e. either in units of the magnetic field $B$ or, for those runs with non-vanishing scalar source $\Lambda$, in units of $\Lambda$).
\section{Numerical solutions with back-reaction}

\label{S:numerics}
At low temperature and sufficiently large Chern-Simons coupling, exact decay rates obtained from solving Einstein-Maxwell-Chern-Simons(-dilaton) equations with back-reaction are expected to differ from and qualitatively alter the results of decay rates estimated from quasi-normal modes solutions, i.e. linearized solutions, which predict modes of infinite life-time in the zero temperature limit. At the very least the temperature increase through the electric quench, creating the modes, will affect the decay rate.  While solving the time-dependent equations with back-reaction at, initially, zero temperature is a challenging  task, solving them at sufficiently low temperature, for long-lived modes to exist (decay rate $\gamma$ of order $10^{-3}$ in units of $\Lambda$ or $\sqrt{B}$) is feasible. \\ \indent
On a Cauchy slice $t=t_i$ we define initial data corresponding to a thermal state in the boundary CFT, with an external magnetic background field, and deformed by an uncharged scalar. We solve the time evolution numerically using the characteristic formulation of GR applied to holography \cite{Chesler:2013lia}. As boundary conditions for the bulk metric and bulk gauge field we choose a flat, Minkowski boundary metric and, next to the static magnetic background field,  an external, time dependent electric field which is smoothly switched on at time $t>t_i$ and smoothly switched off at $t \sim 1/\sqrt{B}$, the source of the scalar operator $\Lambda$ is kept constant. 
For the numerical evolution we define 
\begin{align}
\label{repl_start}
\phi(t,u)&=p_1 u + \frac{1}{3} (p_1^3 + 3 p_1^3 \sigma) u^3 \log(u) + u^2 \psi_0(t, u)\\
a^2_x(t,u)&=a_0(t) + \frac{1}{2} u^2 \log(u) a_0'(t)^2+ u \,(a^2_x)_0(t, u)\\
s(t,u)&=    1 + l u +\frac{1}{18}(-p_1^4 - 3 p_1^4 \sigma) u^4 \log(u)\\
&+ u^2 s_0(t, u)\\
q(t,u)&= 1 + \frac{1}{12} u^4 \log(u) a_0'(t)^2 + u^3 q_0(t, u)\\
A(t,u) &=  \Big(\frac{1}{u}+l\Big)^2 - \frac{p_1^2}{3} + 
  \frac{u^2}{18} \log(u) \big(9 B^2 - 2 p_1^4 \\&- 6 p_1^4 \sigma + 6 a_0'(t)^2\big)+ u \, f(t, u),
  \label{repl_end}
\end{align}
where $l$ is the radial shift parameter which we keep constant during the numerical evolution, as the black brane's apparent horizon is within our integration domain at all times without adjusting $l$.  \\ \indent
\subsection{Solutions with non-vanishing scalar source $\Lambda$}
\label{sec:scalar}
\begin{figure}
    \centering
    \includegraphics[width=0.9\linewidth]{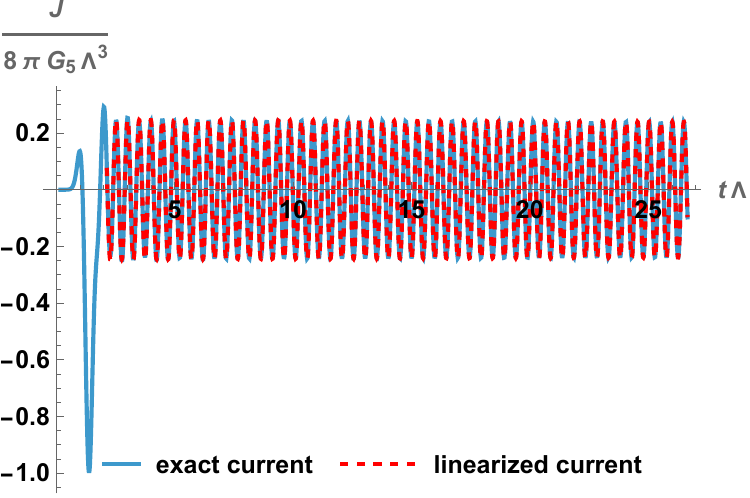}
    \caption{The $x$ component of the current $J=J^2_x$ as a function of time $t\Lambda$, driven out of equilibrium by an electric quench, starting from a magnetic black brane with scalar hair, $B=2.52 \Lambda^2$, and initial temperature $T=0.549 \Lambda$. The scalar-potential-parameter $\sigma$ was set to $1$, such that the potential is bounded. The Chern-Simons coupling $\lambda$ is set to 3 here. The mass squared of the (bulk-) scalar field is $-3$, corresponding to a relevant deformation of the boundary CFT.  We compare the numerically computed results including the full back-reaction of the scalar, the magnetic field and the current (blue solid curve) with the linearized approximation (red dotted curve). We linearize around the geometry at late times. From $t\Lambda =0$ to $t\Lambda \approx 2$ non-linear effects dominate, however, the quasi-normal modes determine the behavior of the current oscillations already from  $t\Lambda \approx 2$ onward. }
    \label{fig:current_b=1.2}
\end{figure}

We numerically construct the initial data corresponding to magnetic black branes with scalar hair via a shooting algorithm. 
We rescale the radial coordinate such that the horizon is positioned at $u=1$ and expand the Einstein equations around the horizon with the ansatz (\ref{ansatz_1}-\ref{ansatz_3}). After fixing values for the scalar-potential-parameter $\sigma$ and the magnetic field $B$ one finds that the near-horizon expansion is unique after fixing four leading order expansion coefficients. Three of these four free parameters correspond to the radial shift freedom  and scaling 
\begin{equation}
    \{x,y\} \rightarrow C_1 \times  \{x,y\}, \,\,z \rightarrow C_2 \times z, \,\, u\rightarrow \frac{u}{1+u\, l},
\end{equation}
the remaining parameter determines the source $p_1 =\Lambda$ of the scalar field after integrating the equations from the horizon to the boundary. We choose $\sigma=1$ and choose the free parameters on the horizon, corresponding to scaling of the spatial coordinates, such that the boundary metric is the Minkowski metric. For Fig. \ref{fig:current_b=1.2}, the initial geometry corresponds to an equilibrium state with scalar source $\Lambda$ with a magnetic background field $B= 2.52\, \Lambda^2$  at (initial) temperature $T_i=0.549 \, \Lambda$. \\ \indent The initially in equilibrium state is driven out of equilibrium by an electric pulse, corresponding to a time dependent source of the bulk gauge field $A^2_x$. For our purpose we choose first
\begin{equation}
\label{quench}
    a_0(t)= \hat{E}_0 e^{-(t-t_1)^2/w^2}
\end{equation}
with $\hat{E}_0 = 0.028 \Lambda$, $t_1\Lambda=1.42$, $w\Lambda=0.284$.
We give a more detailed account of how to numerically solve the fully back-reacted time evolution in \ref{numerics}. 
\begin{figure*}
    \centering
    \includegraphics[width=0.4\linewidth]{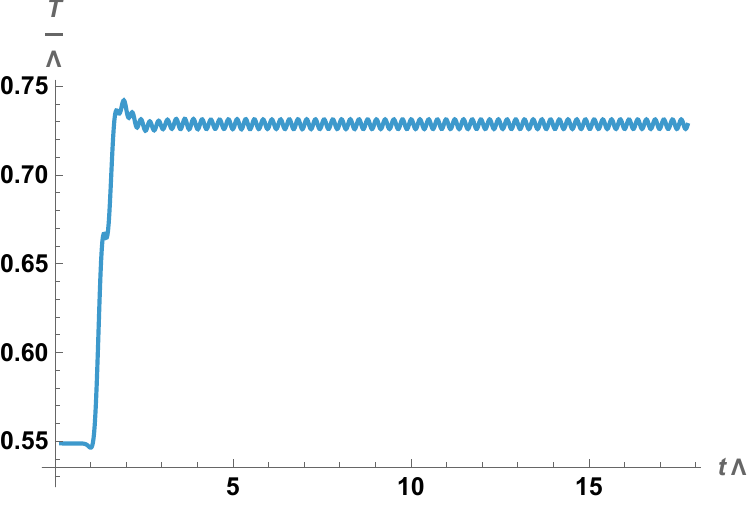}
    \includegraphics[width=0.4\linewidth]{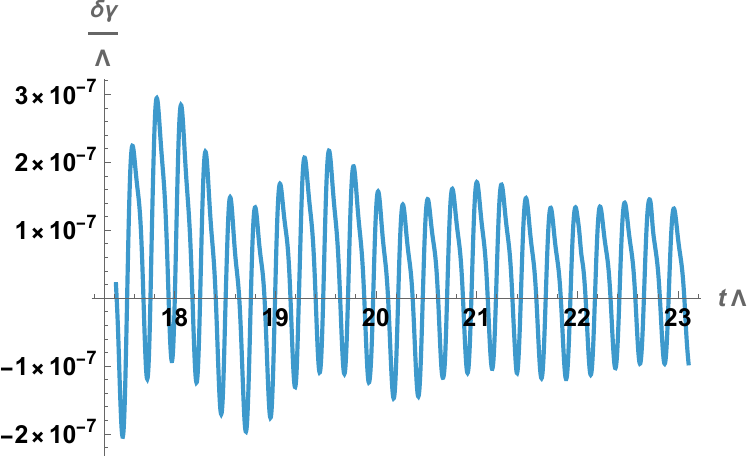}
    \caption{On the left we show the temperature $T/\Lambda$ as a function of time. \HL{ The observable we defined in (\ref{E:temperature}) yields the temperature in equilibrium. Nonetheless, we can use it to show the approximate size of the temperature increase   between early times, where we start in thermal equilibrium, and late times when the geometry is fluctuating around the new equilibrium with twice the frequency of the long lived current modes. Off-equilibrium effects, especially during the rapid growth induced by the electric quench, should not be interpreted as physical predictions for an off-equilibrium or far from equilibrium notion of temperature, as the latter is also not well defined. }. The parameters regarding the potential parameter $\sigma$, the Chern-Simons coupling $\lambda$, the initial temperature, the magnetic background field, the electric quench and the mass of the scalar field are the same as for the evolution in Fig. \ref{fig:current_b=1.2}. On the right we show the difference between the exact and linearized decay rates $\delta \gamma$.}
    \label{fig:temperatureAnddelta}
\end{figure*}
We show the result for the boundary current $J_x^2/(8\pi G_5 \Lambda^3)$ in units of the scalar source $\Lambda$ in figure \ref{fig:current_b=1.2} (note that the upper index in $J_x^2$ is a flavor index, not a squared current). The blue solid curve is the exact, numerically computed result, with full back-reaction of the gauge fields and the scalar field. The red dotted line there is the linearized approximation, where we linearize around the equilibrium state at late times $t\to\infty$. For sufficiently large $\lambda$ and small temperature the decay rate of the mode tends to zero. As also observed in \cite{Meiring:2023wwi} non-linear effects dominate the evolution at early times, but a quasi-normal mode approximation describes the further evolution already early on, here at $t\Lambda \approx 2$. The exact solution is almost indistinguishable from the quasi-normal mode with frequency $\omega/\Lambda =12.8382 - 0.0011 i$, computed for the geometry at late times (we'll quantify the difference in Fig. \ref{fig:temperatureAnddelta}).  We chose the electric pulse in such a way that its Fourier transform's support includes the frequency $\text{Re}(\omega)/\Lambda=12.8382$. \\ \indent As argued in \cite{Waeber:2024ilt}  it is not surprising that we find long lived excitations in the small temperature limit in Einstein-Maxwell-Chern-Simons theory, even after deforming it with a scalar. While qualitative independence of the results from scalar matter content was expected, the results  shown in  the second plot in Fig. \ref{fig:temperatureAnddelta} are surprising. In the first plot on the left hand side in Fig. \ref{fig:temperatureAnddelta}, we show the temperature as a function of time.\HL{ We approximate the temperature $T$ via \begin{equation}
    T(t) \approx \frac{\partial_u A(t,u_0)}{4 \pi}, \text{ with} \,\,\, A(t,u_0)=0.
    \label{E:temperature}
\end{equation}
Strictly speaking, this relation only holds exactly in equilibrium. It yields a good approximation for the temperature initially and at late times, where the persistent current oscillations still induce small metric fluctuations that are imprinted on the observable defined in (\ref{E:temperature}).
}\\ \indent The short electric pulse, creating the current oscillations in Fig. \ref{fig:current_b=1.2}, leads to an increase in temperature, which in turn increases the decay rates, both for exact and linearized solutions. On the right in Fig. \ref{fig:temperatureAnddelta} we show the difference in decay rates 
\begin{equation}
    \delta \gamma(t) = \gamma_{\text{qnm}}(t)-\gamma(t),
    \label{deltagamma}
\end{equation}
where $\gamma(t)$ is the exact decay rate extracted from numerical solutions with back-reaction and $\gamma_{\text{qnm}}(t)$ is the decay rate extracted from quasi-normal mode frequencies  computed around the geometry at time $t$. Expanding the decay rate at late times in the amplitude $\epsilon$ of the current oscillations, once the gauge field source has been switched off, one gets
\begin{equation}
    \gamma = \gamma_{\text{qnm}} +\epsilon^2 \gamma^{(2)}+\mathcal{O}(\epsilon^4).
\end{equation}
The higher order corrections starting at $\epsilon^2$ are a consequence of the field strength entering quadratically as a source to the Einstein equations (\ref{E:EOM}). The anomaly term in the field equations is crucial for the suppression of the imaginary parts of quasi-normal modes, and thus cause $\gamma_{\text{qnm}}$ to approach zero in the zero temperature limit, while the real part of the frequencies remains finite. However, at higher orders, the gauge field fluctuations could excite metric fluctuations for which there is no obvious analogous mechanism to suppress their decay rates, as the dynamical equations for the metric are unaffected by the topological Chern-Simons term. Thus, being fully agnostic about the outcome, one could have suspected that $\gamma_{\text{qnm}} \ll \gamma^{(2)} \approx \mathcal{O}(\sqrt{B}), \mathcal{O}(\Lambda) $ in the small temperature limit, and, thus, that the exact, fully back-reacted solutions' decay rates for the current oscillations are of order $\epsilon^2$. As the right hand side of Fig. \ref{fig:temperatureAnddelta} suggests this does not seem to be the case. In fact, at late times, the difference between the linearized and exact decay rates in units of $\Lambda$ is only of order $10^{-7}$ for this specific choice of parameters, while, as can be derived from Fig. \ref{fig:current_b=1.2} the amplitude squared in units of $\Lambda^6$ of current oscillations is of order $10^{-2}$. As we will also see in the next section, the corrections to the linearized decay rate being at most of order $\epsilon^2 \gamma_{\text{qnm}}$  is more generally the case. This also suggests that exact decay rates become arbitrarily small for sufficiently large $\lambda$ and small temperature $T_{final}$. \\ \indent An explanation of this behavior might be that the $\lambda \to \infty$ limit corresponds to the probe limit, in which the equations of motions for the gauge field fluctuations, dual to the current oscillations, and the equations of motions for the metric plus the scalar field equations, decouple, with the back-reaction of the gauge field fluctuations  becoming negligible. As we will see in the next section \ref{S:results}, the picture does not qualitatively change even for more modest values of $\lambda$, e.g., $\lambda = 1.5$.
  \subsection{Solutions without scalar source $\Lambda=0$ at late times}
\label{S:results}
As in section \ref{sec:scalar} we solve the Einstein-Maxwell-Chern-Simons equation, now with vanishing scalar field, with initial data corresponding to a magnetic black brane in equilibrium, driven by an electric quench. We modify the quench function (\ref{quench}), so that its Fourier transform has a peak at the location at which we expect the real part of the lowest quasi-normal mode, after the quench.  

\begin{figure*}[]
    \centering
    \includegraphics[width=0.45\linewidth]{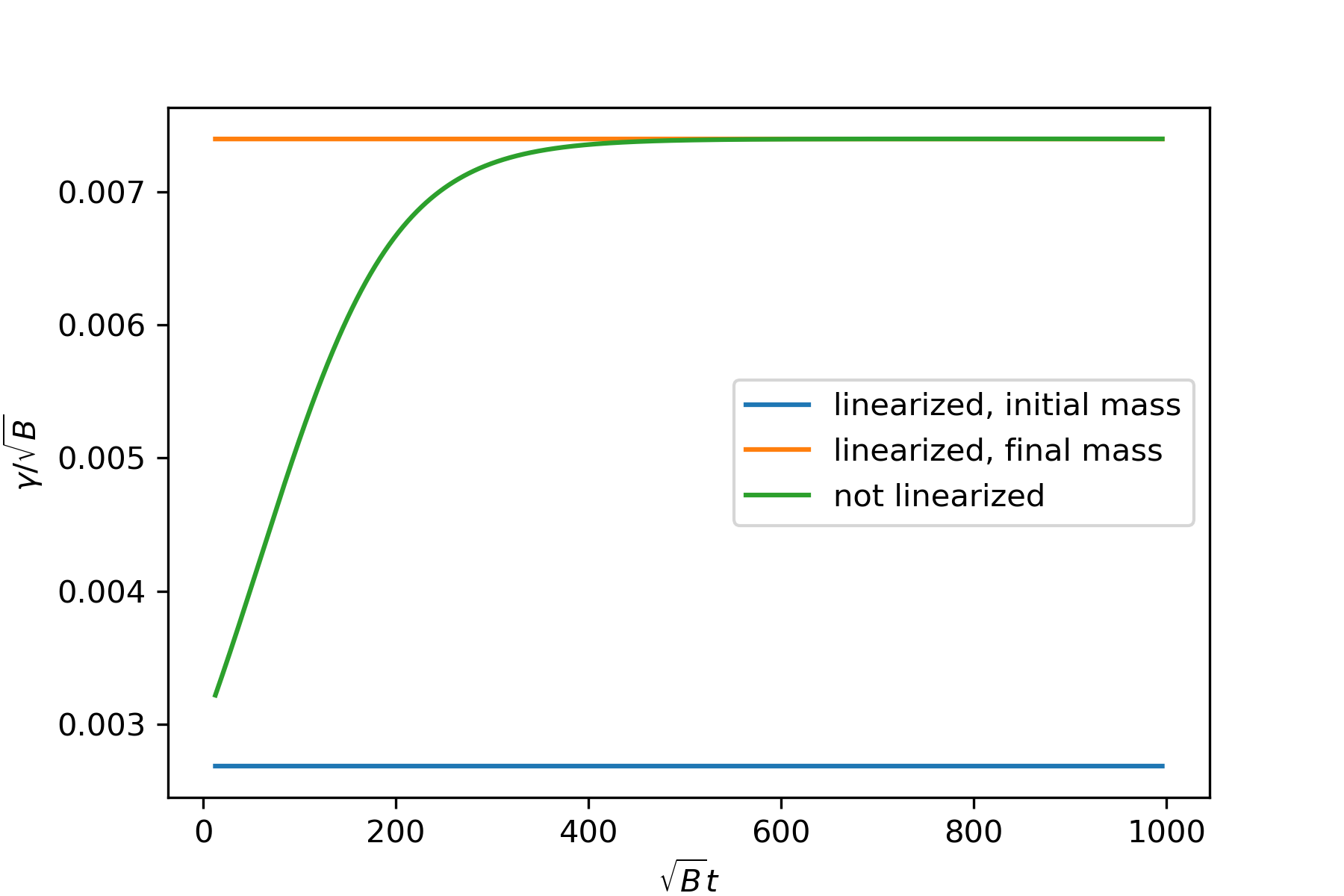}
    \includegraphics[width=0.45\linewidth]{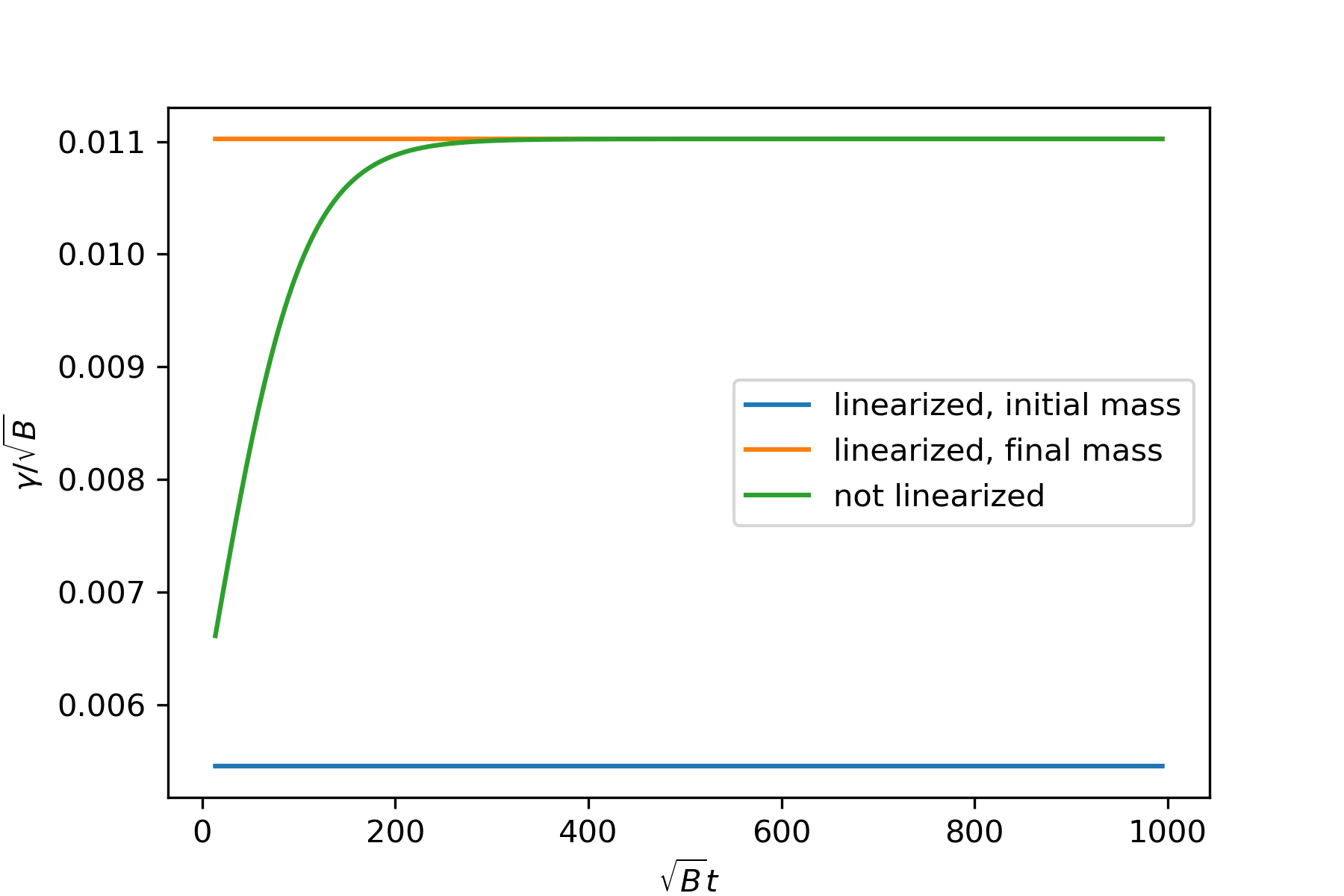}\\
    \includegraphics[width=0.45\linewidth]{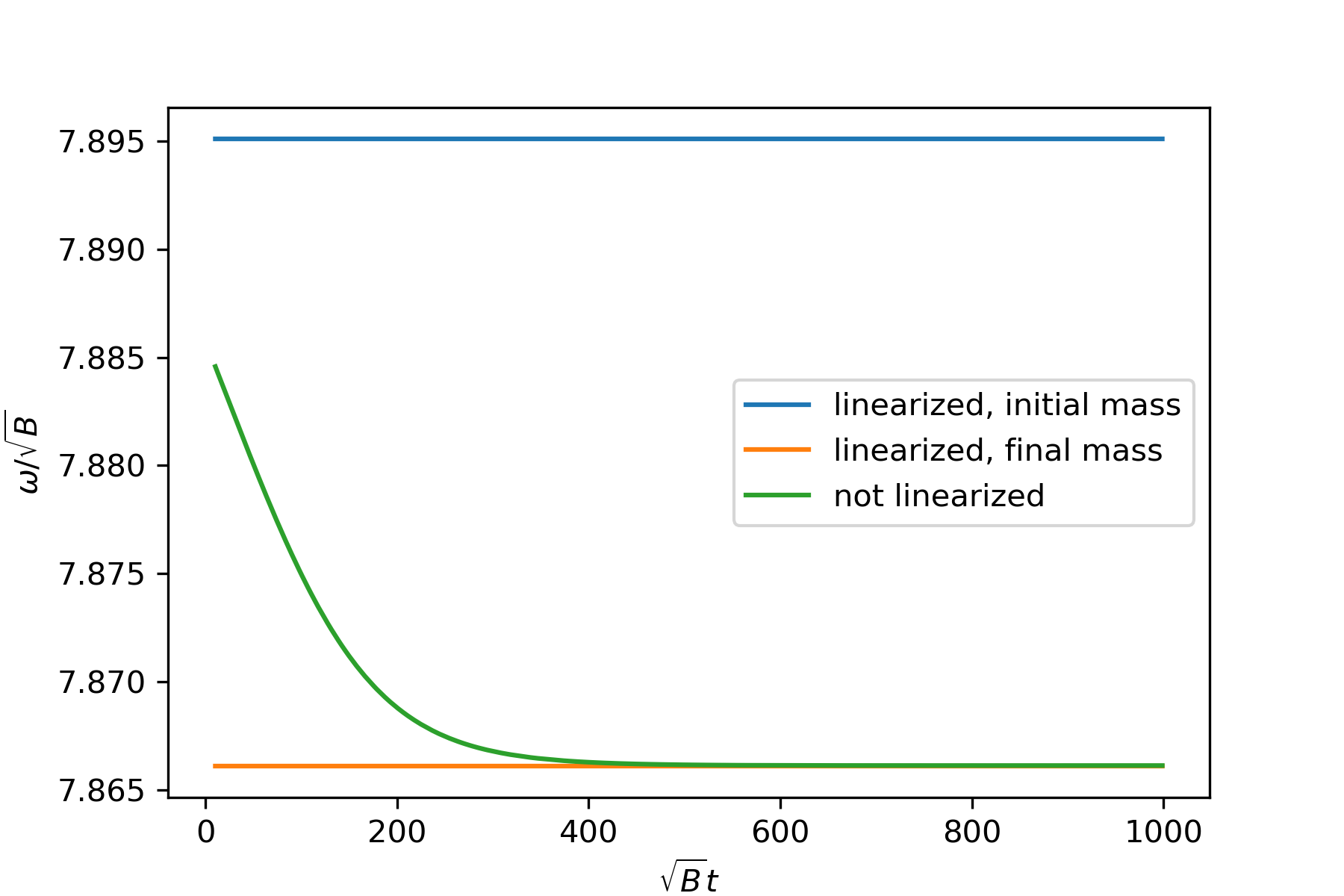}
    \includegraphics[width=0.45\linewidth]{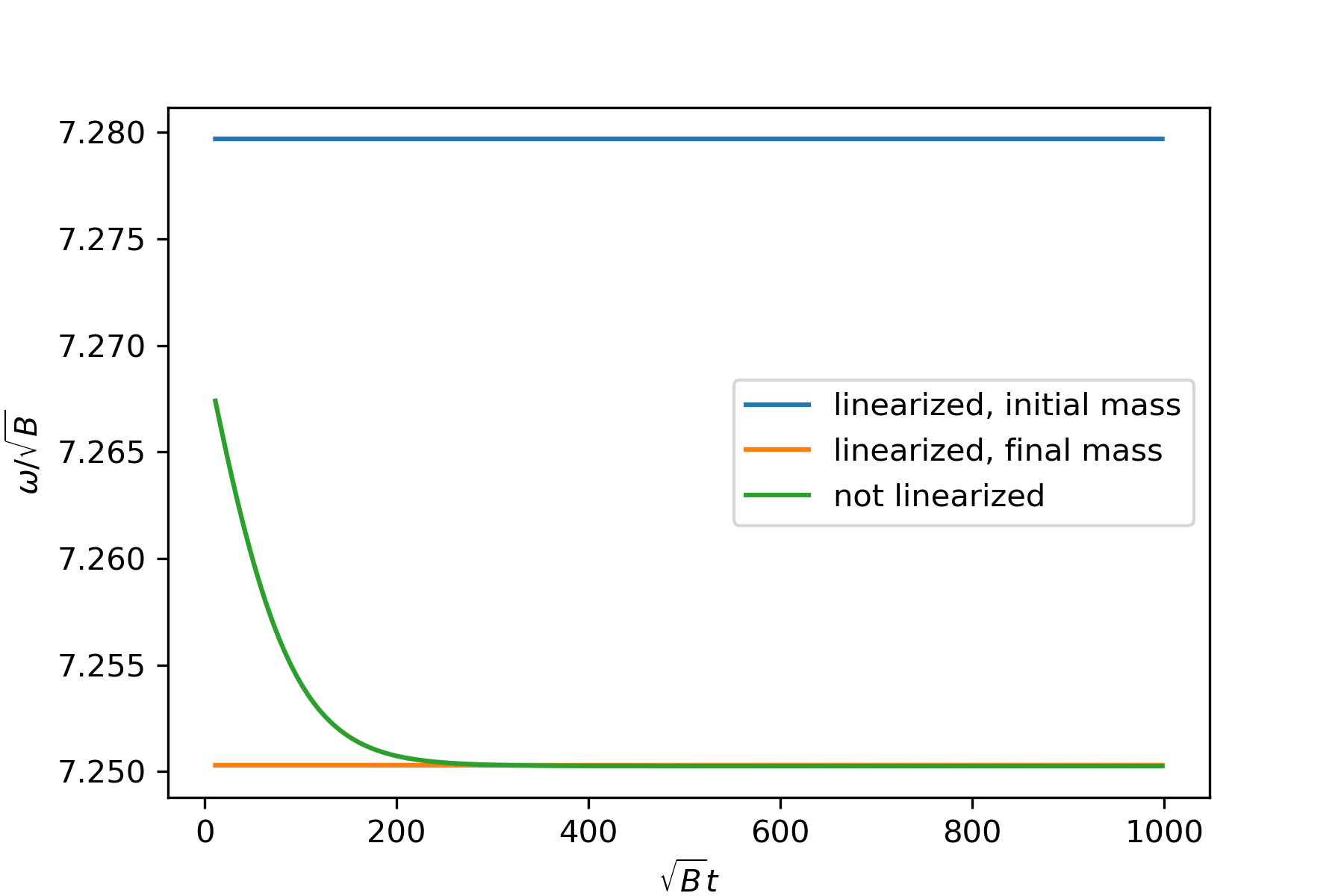}
    \caption{\HL{ The upper two plots show} the linearized and exact decay rates of current fluctuations as a function of time $t\sqrt{B}$. The left side shows results for $\lambda=1.5$, while the right side is for $\lambda=2.5$. The green curves correspond to the time-dependent decay rates calculated from the numerical results with back-reaction. \HL{ We extract the peaks of the current oscillations and fit a linear function to the logarithmic amplitude over a window of ten peaks. The decay rate $\gamma$ is then given by the slope of this linear function.} The blue and orange lines correspond to the linearized decay rates extracted from quasi-normal modes for the initial and final temperature (initial and final mass), respectively. The exact decay rates asymptote towards the linearized decay rates at late times. While this by itself is expected, it is surprising the linearized results approximate the exact results exceedingly well, given the size of the amplitude of the current oscillations (on the left, at, e.g., $t\sqrt{B}=300$, we find that the amplitude $J$ of the current oscillations $J^2_x$ divided by $8\pi G_5$  is $J\approx0.85 B^{3/2} $, on the right,  at $t\sqrt{B}=300$, we find $J\approx0.074 B^{3/2}$). \HL{ The lower two plots show the time-dependent oscillation frequency, compared to the linearized frequencies. The time-dependent frequency is  calculated from the time $\tilde t$ between two peaks of the current oscillation via $\omega=2\pi/\tilde t$. The results are again shown for $\lambda=1.5$ (left) and $\lambda=2.5$ (right). We observe that, in both cases, the oscillation frequency asymptotes to the quasi-normal mode frequency obtained by linearizing around the final geometry.}}
    \label{fig:decay rates}
\end{figure*}

\begin{figure}[h]
    \centering
    \includegraphics[width=\linewidth]{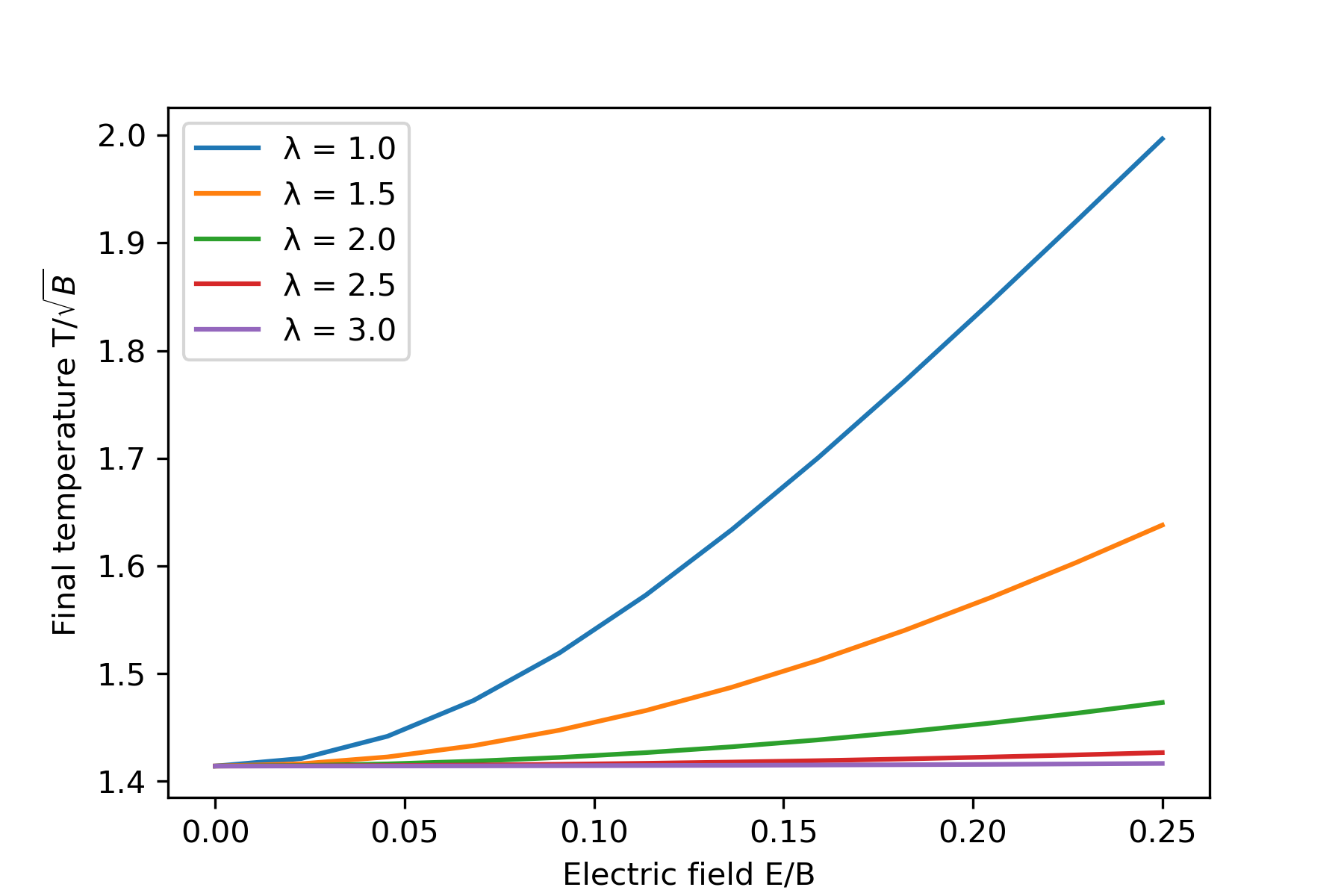}
    \caption{The late time temperature, in units of $\sqrt{B}$, after the electric pulse injected energy, as a function of the amplitude $E$ of the electric pulse in units of $B$. We show the results for various sizes of the Chern-Simons coupling $\lambda$. As the limit $\lambda \to \infty$ corresponds to the probe limit, in which the differential equation for $a^2_x$ and the field equations for the metric decouple, the temperature increase tends to zero for large $\lambda$. Note that the upper index in $a^2_x$ refers to the flavor index of the gauge field, not a squared term.  }
    \label{fig:temp}
\end{figure}

\begin{figure*}
    \centering
    \includegraphics[width=0.47\linewidth]{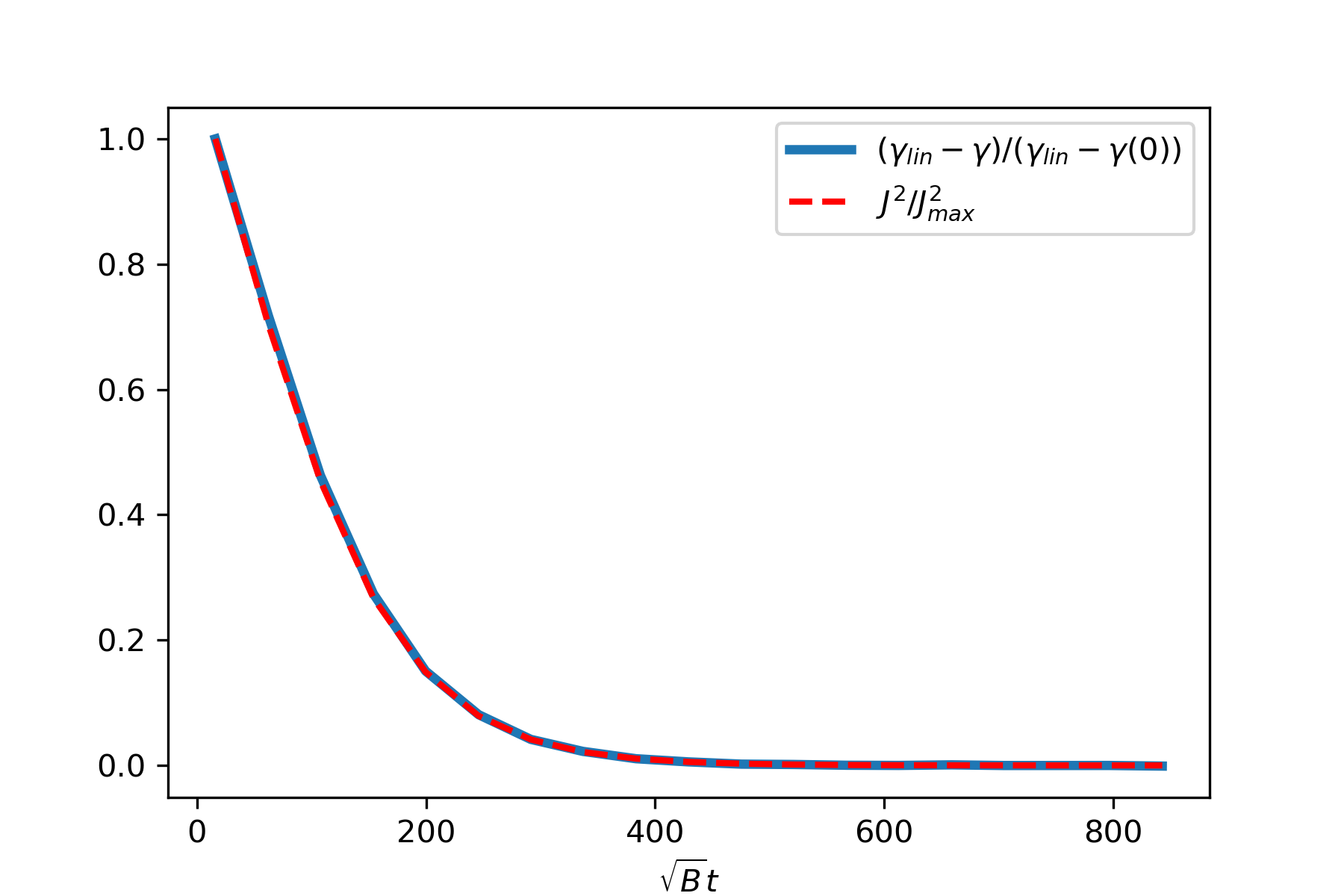}
    \includegraphics[width=0.47\linewidth]{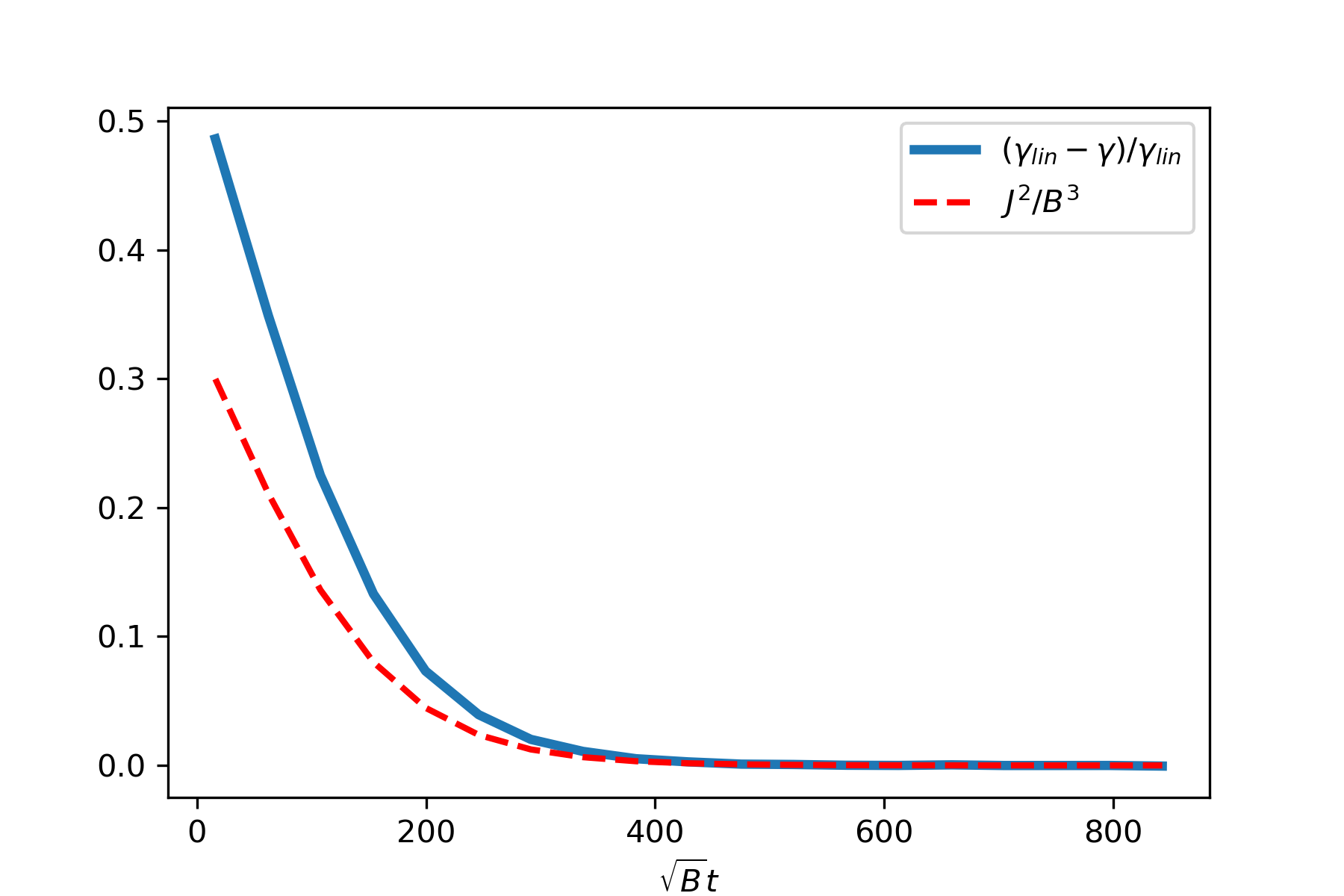}
    \caption{On the left we show the normalized difference between the exact and linearized decay rates (blue curve) compared with the amplitude squared of the current oscillations, normalized by $J_{max}^2$, the maximum of the current's amplitude squared (red curve) after the electric pulse. Both curves agree well. On the right we show the relative difference between exact and linearized decay rates $(\gamma_{\text{lin}}-\gamma)/\gamma_{\text{lin}}$ compared with the amplitude of the current oscillations squared in units of the magnetic field $B$. As can be seen there, the relative difference in decay rates is of the same order. Here $\lambda =1.5$ and the initial temperature is the same as in Fig. \ref{fig:temp}. In contrast to $\delta \gamma =\gamma_{\text{qnm}}-\gamma$, which we defined for Fig. \ref{fig:temperatureAnddelta}, we compare here the exact decay rate $\gamma(t)$ with the linearized decay rate at late times $\gamma_{\text{lin}}=\lim_{t\to \infty}\gamma_{\text{qnm}}$.} \label{fig:decay-rate-difference}
\end{figure*}

We track the residual of the bulk field equations and solve them until late times $t > 800/\sqrt{B}$, for various choices of the Chern-Simons coupling $\lambda$, the initial temperature in units of $\sqrt{B}$ and the amplitude of the electric pulse. From the data for the response of the boundary current to the short-lived electric pulse, perturbing the initial thermal state at $t=0$, we again estimate the decay rate \HL{ by fitting the logarithm of the amplitude of the oscillating current with a linear function. We then compare the slope $\gamma$ of this linear function} to the decay rates obtained from linearized solutions, extracted from the imaginary part of the quasi-normal mode frequencies. We solve the quasi-normal mode equations of the initial geometry, i.e. the magnetic black brane solution corresponding to a thermal state with the initial temperature $T$, in a magnetic background field $B$. These results are shown in Fig. \ref{fig:decay rates}.
At late times, the geometry is fluctuating around an equilibrium solution with increased temperature. The increase in temperature depends on the amplitude of the pulse and the Chern-Simons coupling, see Fig. \ref{fig:temp}. We also solve the linearized Einstein equations and compute  the quasi-normal mode frequencies for the late time geometry. It is not surprising that the imaginary part of the quasi-normal modes of the late time geometry yields a better approximation for the exact decay rates. On the other hand, an unexpected feature of the exact decay rates appears to be that they are bounded from above by the imaginary part of the lowest quasi-normal mode frequency at late times and how well the linearized solutions approximate the exact results. 
We computed the relative difference between the exact decay rates, extracted from the numerical solutions, and the linearized decay rates. We show these results in Fig. \ref{fig:decay-rate-difference}. The normalized difference $(\gamma_{\text{lin}}-\gamma)/(\gamma_{\text{lin}}-\gamma(0))$ between the exact decay rate $\gamma$ and the linearized rate $\gamma_\text{lin}$ matches very well with the normalized current's amplitude squared $J^2/J^2_{\text{max}}$. Here $\gamma_\text{lin}$ is the decay rate extracted from linearizing at late times. So that, comparing it with the definition in (\ref{deltagamma}), $\gamma_\text{lin} =\lim_{t\to\infty}\gamma_{\text{qnm}}$.  This is expected as the back-reaction of the current will affect the decay rates through inducing fluctuations of the metric components sourced by terms quadratic in the bulk gauge fields. On the right in Fig.   \ref{fig:decay-rate-difference} we show the relative difference between the exact and linearized decay rates (linearized around the geometry at late times) $(\gamma_{\text{lin}}-\gamma)/\gamma_{\text{lin}}$, at low temperatures ($T\approx 1.4\sqrt{B}$) and Chern-Simons coupling $\lambda=1.5$, and compare it with the amplitude of the current oscillations squared $J^2$ in units of the magnetic field $B$ (where $J$ is the amplitude of the current $J^2_x$, divided by $8 \pi G_5$). We find that both are of the same order. With $\gamma_{\text{lin}}$ approaching zero in the large $\lambda$ and small temperature limit, this suggests that we can construct approximate, holographic time crystals this way, with arbitrarily small decay rates if $T_{final}$ is sufficiently small and $\lambda$ sufficiently large. 
\section{Discussion}
We numerically solved Einstein-Maxwell-Chern-Simons(-scalar) theory with full back-reaction, starting from a system in equilibrium, driven by time dependent sources for bulk gauge fields. This is dual to the response of a thermal state in a magnetic background field, in a deformed CFT with scalar matter and a anomalously broken $U(1)$ symmetry, to an external electric pulse. As the linearized treatment suggests, at low temperatures and large Chern-Simons coupling $\lambda$, an anomalous resonance effect is observed, with current oscillations decaying slowly. In fact, quasi-normal mode computations at zero temperature \cite{Waeber:2024ilt} yield zero decay rates for $\lambda>1$. The expectation that non-linear effects will increase the decay rate proved to be correct. In the fully back-reacted treatment, the perturbation creating the modes, increases the temperature, which in turn leads to a faster decay of the modes. Surprisingly the exact decay rate does not seem to differ meaningfully from the linearized decay rates, computed from quasi-normal modes at late times, once the temperature no longer increases.  One could have suspected that the size of the exact decay rates in units of the scalar source $\Lambda$ or the magnetic background field $\sqrt{B}$ is of the same order as the amplitude of the current oscillations squared ($\epsilon^2 = \mathcal{O}(J^2/\Lambda^6)$ or, if the scalar source is zero $\epsilon^2 = \mathcal{O}(J^2/B^3)$), as the the bulk field strength enters the Einstein-Maxwell equations quadratically, exciting metric fluctuations of the order $\epsilon^2$. The metric fluctuations are not affected by the topological Chern-Simons term, which in turn is essential to yield quasi-normal mode frequencies for the current oscillations with small imaginary parts $\text{Im}(\omega)\ll \mathcal{O}(\Lambda,\sqrt{B})\lesssim  \text{Re}(\omega) $ at small  temperatures. Thus, our naive expectation was to find the decay rates of the back-reacted solutions to be of order $\epsilon^2$. Instead we found that  the differences between linearized  decay rates (with linearization done for the temperature $T_{final}$ at late times) and exact decay rates are of order $\epsilon^2 \times \text{Im}(\omega)$, where here $\omega$ is a quasi-normal mode of the geometry at late times. For very large $\lambda$, $1\ll \lambda$, this effect is not unexpected as the $\lambda \to \infty$ limit corresponds to the probe limit, so that the back-reaction of the gauge field fluctuations on the geometry becomes negligible. This results in a vanishing effect on the final temperature $T_{final}$ of the electric quench for large $\lambda$. However, we found the exact decay rates to be of order $\epsilon^2 \times \text{Im}(\omega)$  also for modest values of $\lambda$, e.g. $\lambda =1.5$, for which the effects of the gauge field oscillations on the geometry were not negligible.  \\ \indent  While we did not find any indication that this behavior changes at sufficiently small temperature and large Chern-Simons coupling $\lambda$, with the exact decay rates also asymptotically approaching  zero for decreasing temperature and increasing $\lambda$, there is a limit where our numerical evolution becomes unstable, $3 \lesssim \lambda$, $T\lesssim 0.5 \Lambda$. It might be that in the, thus far, numerically inaccessible regime of values for the temperature, the relative difference between the exact and linearized decay rates increases again. \\ \indent There is a variety of future directions into which this research can be expanded in future. They include, studying with full back-reaction, the response of the current to localized injections of energy, tuning the parameters of this system, e.g. potential of the scalar field, to minimize the late time temperature, and maximize the life-time of current oscillations even for smaller values of the Chern-Simons coupling, using machine learning to find, e.g., the optimal parameters for the scalar potential for this purpose, or work on an even more detailed numerical analysis, starting from initial data corresponding to exactly zero temperature solutions. Another interesting question to be tackled in future work is whether we can identify, explicitly in our numerical scheme, the (almost massless) pseudo-Goldstone bosons corresponding to the approximately broken time-translation symmetry in this system with long-lived oscillations. 
\section*{Acknowledgments}
We thank Amos Yarom and Ren\'e Meyer for helpful comments and discussions. SW is grateful for the hospitality of the Perimeter Institute and W\"urzburg University were parts of this work were finalized. 
The work of SW is supported by the Kreitman fellowship program, by the Israel Science Foundation
(grant No. 1417/21), by the German Research Foundation through a German-Israeli Project Cooperation (DIP) grant “Holography and the Swampland”, by Carole and Marcus Weinstein through the BGU Presidential Faculty Recruitment Fund, by the ISF Center of Excellence
for theoretical high energy physics, by the VATAT Research Hub in the Field of quantum computing and by the ERC starting Grant dSHologQI (project number 101117338). MG is supported by the DFG research training group GRK2994.
\begin{appendix}
\section{Solving Einstein-Maxwell-Chern-Simons-dilaton equations}
\label{numerics}
We dynamically solve Einstein-Maxwell-Chern-Simons-dilaton equations starting from a hairy, magnetic black brane in thermal equilibrium,  with magnetic field $B$, a scalar field $\phi$ and scalar source $\Lambda$, which is driven out of equilibrium by an electric quench.    For convenience let us denote the differential equations for the metric (first equation of (\ref{E:EOM})) by $E^{\mu\nu}$ and the Maxwell equations (second equation of (\ref{E:EOM})) by $M^\mu_i$. The constraint equation $M^u_3$ yields the following relation
\begin{equation}
    \partial_t a_t^3 = \frac{8 B \lambda u\,  a^2_x}{s^3},
\end{equation}
which is used to remove $a_t^3 $ from the equations of motion, as only time derivatives of $a_t^3$ enter into dynamical equations. 
The near boundary expansion of the constraint equation $E^{uu}$ yields the energy conservation equation 
\begin{equation}
    \partial_t f_2=(12 a_2 a_0'-6 p_1 p_2-7 a_0' a_0''),
\end{equation} (with $f_2$ the $\mathcal{O}(u^2)$ near boundary expansion coefficient of $A$, defined in (\ref{repl_end}))
whereas the constraint equation $E^{tu}$, the dynamical equations $E^{tt}$, $E^{xx}$, $E^{zz}$, $M^t_2$ and the third equation of (\ref{E:EOM}), are solved on each time slice for the functions 
\begin{equation}
    \{\partial_t \psi_0, \partial_t (a^2_x)_0,\partial_t q_0,f,s_0,\partial_t s_0\}.
\end{equation}
The data needed on each time slice to solve these equations is the radial shift $l$, which we keep constant, as the black hole horizon is contained within our integration domain at all times even without fixing $l$, the near boundary expansion coefficient that determines $\partial_uf(t,0)$ in (\ref{repl_end}), the source $p_1$ which is held constant, as is the magnetic background field $B$, the boundary gauge field $a_0$ (defined in (\ref{repl_start})-(\ref{repl_end})), which is chosen as in (\ref{quench}) for runs with non-zero scalar source, with no spatial dependence and compact support around $t\approx t_1$, where $t_1 \sqrt{B}$ is of order $1$,  and the functions $\psi_0$, $(a_x^2)$, $q_0$ defined in (\ref{repl_start})-(\ref{repl_end}).
On the time slice $t=t_0$ the data 
\begin{multline}
    \text{data} = \{l,p_1,\partial_uf(t_0,0),B,a_0(t_0),\\ \psi(t_0,u),q_0(t_0,u),(a_x^2)(t_0,u)\}
    \label{data}
\end{multline}
is used to solve the Einstein equations which read, with the replacements (\ref{repl_start})-(\ref{repl_end}):
\begin{widetext}
\begin{align}
  0 &=   -\frac{2 u^2 \left((a^2_x)^{(0,1)}(t,u)\right)^2}{q(t,u) \sigma (t,u)^2}-2 \phi^{(0,1)}(t,u)^2-\frac{3 q^{(0,1)}(t,u)^2}{q(t,u)^2}-\frac{6 \sigma ^{(0,2)}(t,u)}{\sigma (t,u)}\\
 0 &= -\frac{1}{4 q(t,u)^2 \sigma (t,u)^6}\Bigg(u^2 \Big(-2 u^2 q(t,u) \sigma (t,u)^4 \left(A(t,u) \Big(u^4 \Big((a^2_x)^{(0,1)}(t,u)\right)^2 \nonumber \\ &-6 u^2 q(t,u) \sigma ^{(0,1)}(t,u)^2\Big)+12 q(t,u) \sigma ^{(0,1)}(t,u) \sigma ^{(1,0)}(t,u)\Big)+64 B^2 \lambda ^2 u^6 q(t,u)^2 \nonumber \\ &
   ((a^2_x)(t,u))^2+\sigma (t,u)^6 \Big(2 q(t,u)^2 \Big(-u^2 \Big(3 u A^{(0,1)}(t,u)+A(t,u) \Big(u^2 \phi^{(0,1)}(t,u)^2 \nonumber \\ &-6\Big)\Big)-3 \phi(t,u)^2+\sigma  \phi(t,u)^4-12\Big)-3 u^4 A(t,u) q^{(0,1)}(t,u)^2\Big)+6 u
   q(t,u)^2 \nonumber \\ &\sigma (t,u)^5 \Big(u \Big(u \Big(u A^{(0,1)}(t,u)-4 A(t,u)\Big) \sigma ^{(0,1)}(t,u)-2 \sigma ^{(1,1)}(t,u)\Big)+6 \sigma ^{(1,0)}(t,u)\Big)\nonumber \\ &+B^2 u^4 \Big(2 q(t,u)^3+1\Big) \sigma (t,u)^2\Big)\Bigg) 
   \end{align}
\begin{align}
   0&=\frac{1}{4 q(t,u)^3 \sigma (t,u)^8}\Bigg(u^2 \Big(-64 B^2 \lambda^2 u^6 q(t,u)^2 ((a^2_x)(t,u))^2+\sigma (t,u)^6 \Big(2 q(t,u)^2 \Big(u^2 \Big(u^2 \nonumber \\ & A^{(0,2)}(t,u)-2 u A^{(0,1)}(t,u)+A(t,u) \Big(u^2 \phi^{(0,1)}(t,u)^2+2\Big)-2 \phi^{(0,1)}(t,u)
   \phi^{(1,0)}(t,u)\Big)+\nonumber \\ &\sigma  \phi(t,u)^4-3 \phi(t,u)^2-12\Big)-2 u q(t,u) \Big(u \Big(u \Big(\Big(u A^{(0,1)}(t,u)-A(t,u)\Big) q^{(0,1)}(t,u)+\nonumber \\ &u A(t,u) q^{(0,2)}(t,u)\Big)-2 q^{(1,1)}(t,u)\Big)+3
   q^{(1,0)}(t,u)\Big)+5 u^2 q^{(0,1)}(t,u) \Big(u^2 A(t,u) q^{(0,1)}(t,u)\nonumber \\ &-2 q^{(1,0)}(t,u)\Big)\Big)+2 u q(t,u) \sigma (t,u)^5 \Big(4 q(t,u) \Big(u^3 A^{(0,1)}(t,u) \sigma ^{(0,1)}(t,u)-2 u \sigma ^{(1,1)}(t,u)+\nonumber \\ &3 \sigma
   ^{(1,0)}(t,u)\Big)+u^2 A(t,u) \Big(4 u q(t,u) \sigma ^{(0,2)}(t,u)-\Big(3 u q^{(0,1)}(t,u)+4 q(t,u)\Big) \sigma ^{(0,1)}(t,u)\Big)\nonumber \\ &+3 u \Big(q^{(1,0)}(t,u) \sigma ^{(0,1)}(t,u)+q^{(0,1)}(t,u) \sigma ^{(1,0)}(t,u)\Big)\Big)+4 u^2
   q(t,u)^2 \sigma (t,u)^4 \sigma ^{(0,1)}(t,u)\nonumber \\ & \Big(u^2 A(t,u) \sigma ^{(0,1)}(t,u)-2 \sigma ^{(1,0)}(t,u)\Big)-B^2 u^4 \sigma (t,u)^2\Big)\Bigg)
  \end{align}
\begin{align}
   0&=u^2 \Big(2 u^2 q(t,u) \sigma (t,u)^4 \Big(A(t,u) \Big(u^4 \Big((a^2_x)^{(0,1)}(t,u)\Big)^2+2 u^2 q(t,u) \sigma ^{(0,1)}(t,u)^2\Big)-\nonumber \\ &2 \Big(u^2 (a^2_x)^{(0,1)}(t,u) (a^2_x)^{(1,0)}(t,u)+2 q(t,u) \sigma ^{(0,1)}(t,u) \sigma
   ^{(1,0)}(t,u)\Big)\Big)-64 B^2 \lambda ^2 u^6 q(t,u)^2 \nonumber \\ &((a^2_x)(t,u))^2+\sigma (t,u)^6 \Big(2 q(t,u)^2 \Big(u^2 \Big(u^2 A^{(0,2)}(t,u)-2 u A^{(0,1)}(t,u)+A(t,u) \Big(u^2 \nonumber \\ &\phi^{(0,1)}(t,u)^2+2\Big)-2 \phi^{(0,1)}(t,u)
   \phi^{(1,0)}(t,u)\Big)+\sigma  \phi(t,u)^4-3 \phi(t,u)^2-12\Big)+\nonumber \\ &4 u q(t,u) \Big(u \Big(u \Big(\Big(u A^{(0,1)}(t,u)-A(t,u)\Big) q^{(0,1)}(t,u)+u A(t,u) q^{(0,2)}(t,u)\Big)-2 q^{(1,1)}(t,u)\Big)\nonumber \\ &+3
   q^{(1,0)}(t,u)\Big)+u^2 q^{(0,1)}(t,u) \Big(2 q^{(1,0)}(t,u)-u^2 A(t,u) q^{(0,1)}(t,u)\Big)\Big)+4 u q(t,u) \sigma (t,u)^5\nonumber \\ & \Big(2 q(t,u) \Big(u^3 A^{(0,1)}(t,u) \sigma ^{(0,1)}(t,u)-2 u \sigma ^{(1,1)}(t,u)+3 \sigma
   ^{(1,0)}(t,u)\Big)+u^2 A(t,u) \Big(\Big(3 u \nonumber \\ &q^{(0,1)}(t,u)-2 q(t,u)\Big) \sigma ^{(0,1)}(t,u)+2 u q(t,u) \sigma ^{(0,2)}(t,u)\Big)-3 u \Big(q^{(1,0)}(t,u) \sigma ^{(0,1)}(t,u)+\nonumber \\ &q^{(0,1)}(t,u) \sigma ^{(1,0)}(t,u)\Big)\Big)+B^2 u^4
   \Big(1-2 q(t,u)^3\Big) \sigma (t,u)^2\Bigg)
\end{align}
and scalar and gauge field equations
\begin{align}
    0&= \frac{1}{\sigma (t,u)}\Bigg(u \Big(-\sigma (t,u) \Big(u^3 A^{(0,1)}(t,u) \phi^{(0,1)}(t,u)-2 u \phi^{(1,1)}(t,u)+3 \phi^{(1,0)}(t,u)\Big)+\nonumber \\ &u^2 A(t,u) \Big(\left(\phi^{(0,1)}(t,u)-u \phi^{(0,2)}(t,u)\right) \sigma (t,u)-3 u
   \phi^{(0,1)}(t,u) \sigma ^{(0,1)}(t,u)\Big)\nonumber \\ &+3 u \left(\phi^{(1,0)}(t,u) \sigma ^{(0,1)}(t,u)+\phi^{(0,1)}(t,u) \sigma ^{(1,0)}(t,u)\right)\Big)\Bigg)+2 \sigma  \phi(t,u)^3-3 \phi(t,u)\\
   0 & = u^3 \Big(-q(t,u) \sigma (t,u)^3 \Big(\sigma (t,u) \Big(u^3 A^{(0,1)}(t,u) (a^2_x)^{(0,1)}(t,u)-2 u (a^2_x)^{(1,1)}(t,u)+\nonumber \\ &(a^2_x)^{(1,0)}(t,u)\Big)+u^2 A(t,u) \Big(\left(u \sigma ^{(0,1)}(t,u)+\sigma (t,u)\right) (a^2_x)^{(0,1)}(t,u)+u \sigma (t,u)
   (a^2_x)^{(0,2)}(t,u)\Big)\nonumber \\ &-u \left(\sigma ^{(1,0)}(t,u) (a^2_x)^{(0,1)}(t,u)+\sigma ^{(0,1)}(t,u) (a^2_x)^{(1,0)}(t,u)\right)\Big)+u \sigma (t,u)^4 \Big((a^2_x)^{(0,1)}(t,u)\nonumber \\ & \left(u^2 A(t,u) q^{(0,1)}(t,u)-q^{(1,0)}(t,u)\Big)-q^{(0,1)}(t,u)
   a(1,2)^{(1,0)}(t,u)\right)\nonumber \\ &+32 B^2 \lambda ^2 u^3 q(t,u)^2 (a^2_x)(t,u)\Big).
\end{align}
The solutions to these equations together with the constraint $E^{uu}$ yields then time derivatives of all time dependent functions in (\ref{data}). We then use a  Runge-Kutta algorithm (RK4), to integrate the equations in time. 
\end{widetext}

\end{appendix}

\bibliographystyle{apsrev4-2}
\bibliography{LLM}

\end{document}